\documentclass[iop,apj,twocolumn]{emulateapj}
\usepackage{natbib,graphicx}
\usepackage{epstopdf}                                      
\usepackage[breaklinks,colorlinks,urlcolor=blue,citecolor=blue,linkcolor=blue]{hyperref} 
\newcommand{\teff}{T_{\mathrm{eff}}}

\begin{document}

\title{Kepler Monitoring of an L Dwarf II. Clouds with Multiyear Lifetimes}

\author{John E.\ Gizis\altaffilmark{1}$^,$\altaffilmark{2}, 
Kyle G.\ Dettman\altaffilmark{1}, 
Adam J.\ Burgasser\altaffilmark{3},
Sara Camnasio\altaffilmark{4}$^,$\altaffilmark{2},
Munazza Alam\altaffilmark{4}$^,$\altaffilmark{2},
Joseph C. Filippazzo\altaffilmark{5}$^,$\altaffilmark{6}$^,$\altaffilmark{2},
Kelle L.\ Cruz\altaffilmark{4}$^,$\altaffilmark{5} ,
Stanimir Metchev\altaffilmark{7},
Edo Berger\altaffilmark{8}, and
Peter K.\ G.\ Williams\altaffilmark{8}}

\altaffiltext{1}{Department of Physics and Astronomy, University of Delaware, Newark, DE 19716, USA}
\altaffiltext{2}{Visiting astronomer, Kitt Peak National Observatory, National Optical Astronomy Observatory, which is operated by the Association of Universities for Research in Astronomy (AURA) under a cooperative agreement with the National Science Foundation. }
\altaffiltext{3}{Center for Astrophysics and Space Science, University of California San Diego, La Jolla, CA 92093, USA}
\altaffiltext{4}{Department of Physics and Astronomy, Hunter College, City University of New York, 695 Park Avenue, New York, NY 10065, USA}
\altaffiltext{5}{Department of Astrophysics, American Museum of Natural History, Central Park West at 79th Street, New York, NY 10034, USA} 
\altaffiltext{6}{Department of Engineering Science and Physics, College of Staten Island, 2800 Victory Boulevard, Staten Island, NY 10301, USA}
\altaffiltext{7}{The University of Western Ontario, Department of Physics and Astronomy, 1151 Richmond Avenue, London, ON N6A 3K7, Canada}
\altaffiltext{8}{Harvard-Smithsonian Center for Astrophysics, 60 Garden Street, Cambridge, MA 02138, USA}

\begin{abstract}
We present Kepler, Spitzer Space Telescope, Gemini-North, MMT, and Kitt Peak observations of the L1 dwarf WISEP J190648.47+401106.8.  We find that the Kepler optical light curve is consistent in phase and amplitude over the nearly two years of monitoring with a peak-to-peak amplitude of 1.4\%. Spitzer Infrared Array Camera 3.6 \micron~observations are in phase with Kepler with similar light curve shape and peak-to-peak amplitude 1.1\%, but at 4.5 \micron, the variability has amplitude $<0.1$\%.  Chromospheric H$\alpha$ emission is variable but not synced with the stable Kepler light curve.  A single dark spot can reproduce the light curve but is not a unique solution. An inhomogeneous cloud deck, specifically a region of thick cloud cover, can explain the multi-wavelength data of this ultracool dwarf and need not be coupled with the asynchronous magnetic emission variations. The long life of the cloud is in contrast with weather changes seen in cooler brown dwarfs on the timescale of hours and days. 
\end{abstract}


\section{Introduction\label{intro}}

The condensation of minerals at temperatures below 2500K shapes the properties of very-low-mass stars and brown dwarfs (see the reviews of \citealt{1997ARA&A..35..137A}, \citealt{2001RvMP...73..719B}, \citealt{2013cctp.book..367M}).
Condensation depletes the gas phase of molecules, but the grains can collect into clouds, removing some sources of opacity and adding others.  The clouds not only change the temperature-pressure profile of the atmosphere, but by changing the boundary conditions of the fully-convective interior, lower the effective temperature and mass of the hydrogen-burning limit into the early-L spectral type. The observationally-defined spectral type sequence from M8 and M9 dwarfs to L dwarfs to T dwarfs is understood as a temperature sequence, with clouds near the photosphere for L dwarfs and below it for T dwarfs (see the review of \citealt{2005ARA&A..43..195K}).  

Clouds immediately suggest the possibility of inhomogeneous coverage and rotational modulation, and searches for variability \citep{Tinney:1999fk,1999A&A...348..800B} began even as the L dwarf spectral type system was being defined. Optical searches \citep{2001A&A...367..218B,2002MNRAS.335.1158C,2002MNRAS.332..361C,2003MNRAS.341..239C,Gelino:2002uq,2003MNRAS.346..473K,2004MNRAS.354..378K,2005MNRAS.360.1132K,2005MNRAS.357.1151K,2006MNRAS.367.1735K,2013MNRAS.428.2824K,2005ApJ...619L.183M,2007AJ....133.1633M,Lane:2007yq} have often found evidence of variability in early L dwarfs. Recently, the ``Weather on Other Worlds" survey \citep{2015ApJ...799..154M} using the Spitzer Space Telescope found that spots are ubiquitous in L and T dwarfs, with a detailed study of the L3 dwarf DENIS-P J1058.7-1548 demonstrating the need for inhomogeneous clouds  to explain the observations \citep{Heinze:2013uq}.
Near-infrared studies have also detected variability in L and T spectral types \citep{Khandrika:2013fk,2014A&A...566A.111W,2014ApJ...782...77B,2014ApJ...793...75R}.  Remarkably, a number of very high-amplitude early-T dwarf variables have patchy clouds that change on the timescale of hours and days \citep{Artigau:2009ly,Radigan:2012vn,2013A&A...555L...5G,2014ApJ...785...48B}. \citet{2013ApJ...768..121A} use multi-wavelength near-infrared observations to show that early T dwarfs are consistent with distinct regions of thin and thick clouds.  \citet{2012ApJ...760L..31B} used simultaneous Spitzer and Hubble observations of a variable T6.5 dwarf to show pressure-dependent phase shifts, taken to be evidence of both vertical and horizontal structure in the clouds. The long-term evolution of clouds, however, is poorly understood because it is usually impossible to observe more than a few rotation periods.  The Kepler Mission \citep{2010ApJ...713L..79K} was  designed to measure photometry for long, continuous time series.  The L1 dwarf \objectname[WISEP J190648.47+401106.8]{WISEP J190648.47+401106.8} \citep{Gizis:2011lr}, hereafter W1906+40, happens to lie in the Kepler field of view, and we observed it in Director's Discretionary Time (GO30101) and Guest Observer (GO 40004) programs.   

In \citet{2013ApJ...779..172G} (hereafter Paper I), we reported initial Kepler results on W1906+40, which has the spectrum of a typical L1 dwarf, probably on the stellar side of the hydrogen-burning limit and a trigonometric parallax distance of $16.35^{+0.36}_{-0.34}$ pc. Over the first 15 months of Kepler observations, it was variable at a 1.4\% level with a period of 8.9 hours and a stable light curve. The Kepler filter is very broad, from 430 nm to 900 nm, but because W1906+40 is very red, the effective wavelength is 800nm.  W1906+40 was also found to be magnetically active, with quiescent radio and H$\alpha$ emission and transient white light flares (Paper I) . Because Kepler has only one filter, we had no information on the spectral characteristics of the spot. In this paper, we analyze the complete Kepler dataset  and use supporting ground-based and Spitzer observations to characterize the nature of W1906+40's variability.

\section{Data and Observations\label{sec-data}}

\subsection{Kepler Photometry}

W1906+40 was observed during Kepler Quarters 10-17 in long-cadence mode \citep{Jenkins:2010fk}, providing photometric measurements every 30 minutes for nearly two years until the end of the original mission. Here we re-analyze the Quarters 10-14 data reported in Paper I, and add the data collected in Quarters 15-17.   A new Kepler Input Catalog (KIC) identification for W1906+40 was generated each quarter (Table~\ref{tab:ids}). Because the spacecraft was rotated by 90 degrees quarterly, each position was observed twice (i.e., W1906+40 is on the same CCD pixels during Quarters 10 and 14).  We use the Kepler pixel-response function 
\citep{2010ApJ...713L..97B} fitting (PRF) routines in the PyKE package \citep{2012ascl.soft08004S} to measure photometry.  This weights the pixels more optimally than simple aperture photometry and allows us to include a model  for potential contamination from the background star KIC 4996079, though it turns out to be negligible. 
The median count rates are given in Table~\ref{tab:ids}; the variation by quarter is instrumental. Each quarter's photometry is normalized so that the median Kepler count rate is 1.0; in these units the uncertainty of each measurement is 0.005.  We remove instrumental long-term trends with low-order polynomials; this removes the known instrumental variations on the timescales of weeks to months (see the discussion in \citealt{Stumpe:2012uq}) but does not affect shorter timescales. The 8.9 hour periodic variability is persistent throughout the observations. We examined all the data by eye and find no convincing episodes where the variability disappeared.

\begin{deluxetable}{ccc}
\tablewidth{0pc}
\tabletypesize{\footnotesize}
\tablenum{1}
\tablecaption{Kepler Identifiers for WISEP J190648.47+401106.8 }
\tablehead{
\colhead{Quarter} &  
\colhead{KIC} &
\colhead{PRF Flux (e$^{-}$/s)}
}
\startdata
10 & 100003560 & 295 \\
11 & 100003605 & 343 \\
12 & 100003905 & 357 \\
13 & 100004035 & 256 \\
14 & 100004076 & 302  \\
15 & 100004142 & 343 \\
16 & 100004180 & 360 \\
17 & 100004298 & 255 \\
\enddata
\label{tab:ids}
\end{deluxetable}

\begin{figure*}
\plotone{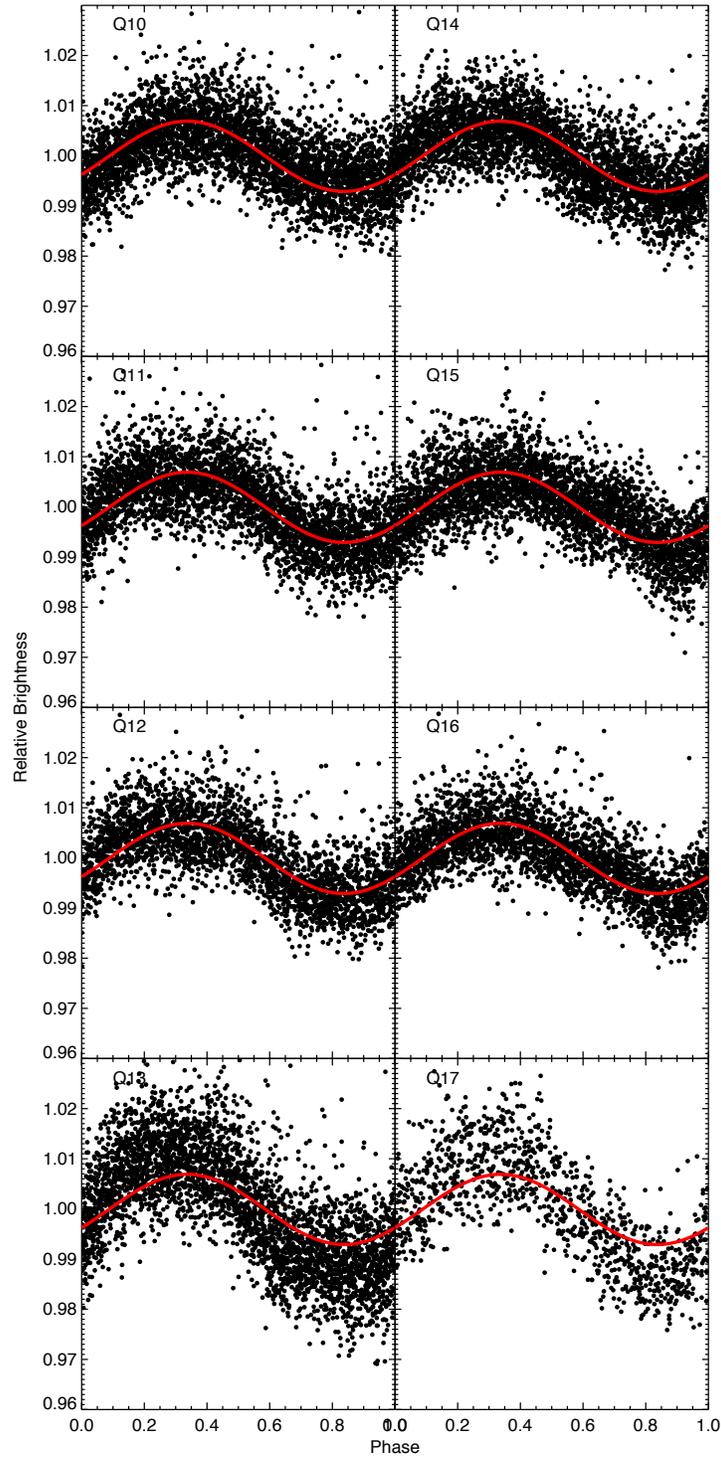}
\caption{Observed Kepler data for each quarter. A sine curve with period of 0.3710177 days and peak-to-peak amplitude 1.4\% is also plotted. Note the consistent phase for all quarters. See text for discussion of the systematic error in Quarters 13 and 17. \label{fig-keplerall}}
\end{figure*}

\begin{figure}
\plotone{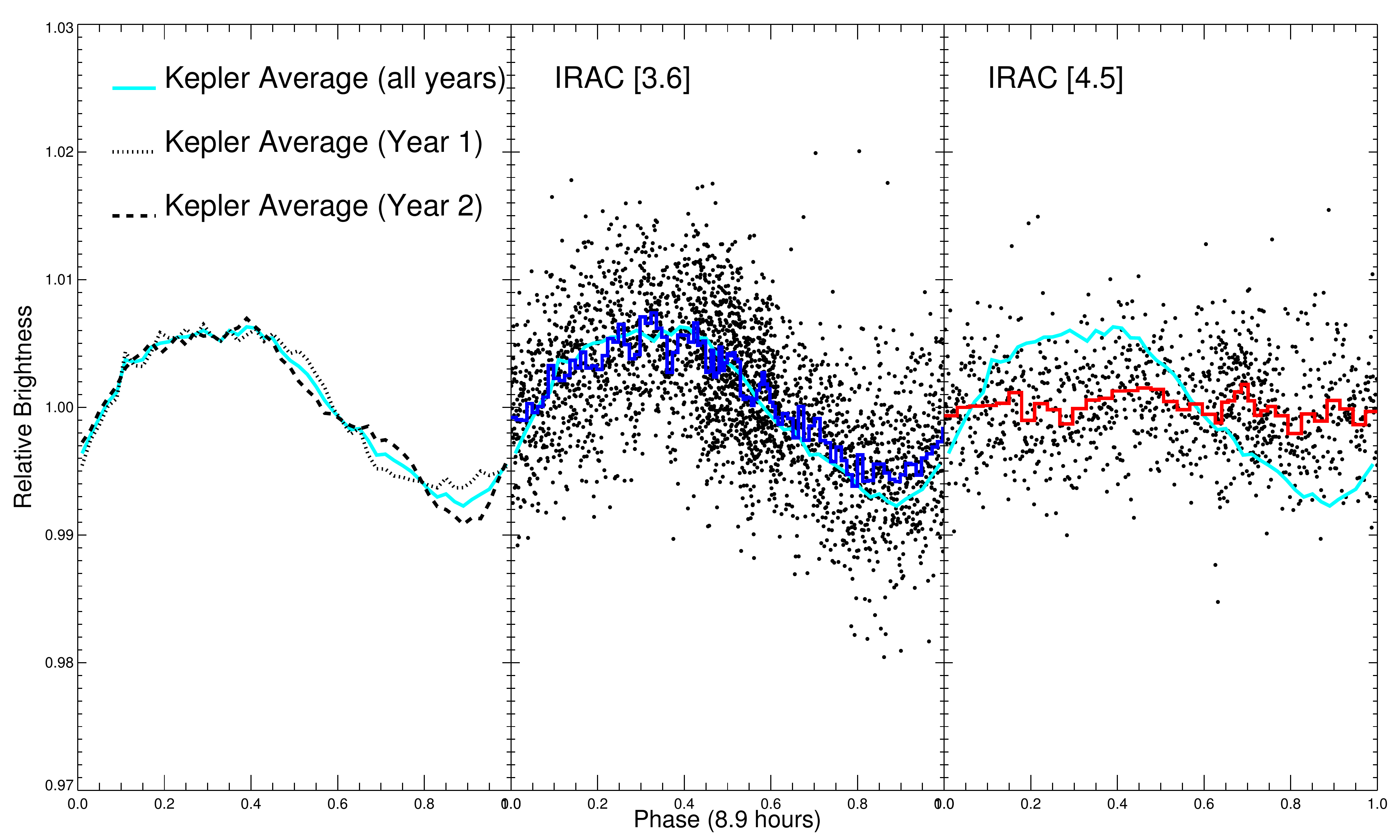}
\caption{Two-dimensional histogram of Kepler brightnesses for W1906+40 in Quarters 10-12 and 14-16 phased to the common period of 0.3710177 days. \label{fig-keplerhist}}
\end{figure}

The Quarters 13 and 17 data show systematic differences from the other quarters; the period and phase is the same, but average count rate is lower and the percentage amplitude of the variability is higher. As noted in Paper I, two of the pixels yield negative counts and must be excluded from the aperture photometry. We obtain similar light curves for either aperture or PRF photometry. Since the different light curve is correlated with the detector used, we attribute it to an instrumental effect; a possible explanation is that the pipeline sky subtraction is incorrect.  Quarter 10 begins at Kepler mission day 906.87, Quarter 16 ends at Kepler mission day 1558.0, and Quarter 17 ends at day 1591.0.\footnote{Kepler mission days are barycentric julian dates (BJD) minus 2454833.0.} We exclude Quarters 13 and 17 from further analysis. On the basis of the single spot model fitting described later, we adopt a period of 0.370177 days with phase zero defined to be Kepler Mission time 1180.0, near the midpoint of our observations. (This is a different time for phase zero than in Paper I). In Figure~\ref{fig-keplerhist}, we show the two-dimensional histogram of the Quarters 10-12 and 14-16 data. We compute binned Kepler average light curves shown in Figure~\ref{fig-spitzerkepler}.  Small differences between the first year and second year observations are evident. The uncertainty of each measurement is a third of the peak-to-peak amplitude, so we can only detect such subtle changes in the light curve by combining data over long time periods.

\subsection{Spitzer Photometry}

\begin{figure*}
\plotone{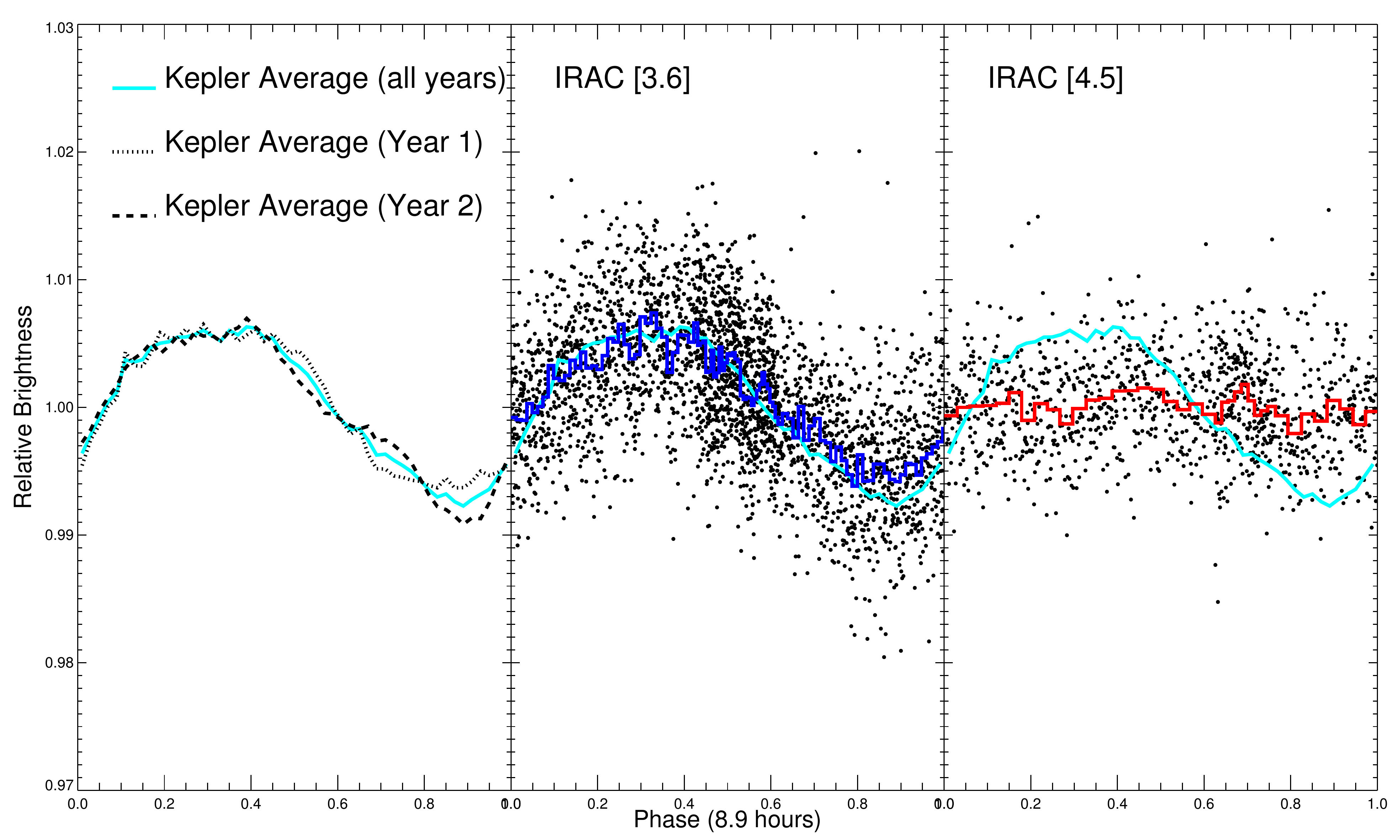}
\caption{Phased light curves for a period of 0.370177 days. The average Kepler light curve is in cyan in all panels. Left: Kepler light curve for first year [Quarters 10-12, days 907-1182] (dotted) and second year [Quarters 14-16, days 1274-1558] (dashed).  Middle: Spitzer IRAC [3.6]  measurements (points) and averaged values (blue histogram) on Kepler Mission day 1750.  Right: Spitzer IRAC [4.5] measurements (points) and averaged values (red histogram). We estimate that the peak-to-peak amplitude of variabilitiy is 1.4\% at Kepler, 1.1\% at [3.6], and 0.04\% at [4.5] (See Section~\ref{sec:spot}). \label{fig-spitzerkepler}}
\end{figure*}

W1906+40 was observed (Program ID 90152) with the Spitzer Space Telescope  \citep{2004ApJS..154....1W} using the Infrared Array Camera (IRAC, \citealt{2004ApJS..154...10F}) for twenty hours on UT Dates 17-18 October 2013, corresponding to Kepler Mission day 1750. This is after the end of the Kepler monitoring on day 1591. The target was located at pixel positions X=23.5, Y=231, near the ``sweet spot" recommended by the Spitzer Science Center, with no dithering in order to minimize the pixel-phase correction \citep{2012SPIE.8442E..1YI}. The observations consist of 130 12-second exposures in IRAC Channel 1 (hereafter [3.6])  while the telescope pointing settled, 2910 12-second exposures in [3.6] used for science, and then 1164 30-second exposures in IRAC Channel 2 (hereafter [4.5]).  For relative photometry, we use 3-pixel aperture photometry with a 12-20 pixel sky annulus. For comparison to other L dwarfs, we apply the Vega zero-points, aperture corrections and array-location dependent corrections provided by the Spitzer Science Center to obtain median magnitudes of 11.24 and 11.23 at [3.6] and [4.5] respectively. The color is consistent with other early L dwarfs measured by \citet{2006ApJ...651..502P}: L dwarfs have near-zero [3.6]-[4.5] colors like A0 stars despite their lower temperatures because of the strong CO absorption in the [4.5] band.  Our observations show a slow drift in position over the entire observation plus an instrumental periodic ($\sim 51$ minute) pattern; the total change in position over 0.2 pixels in X and 0.3 pixels in Y during the [3.6] observations and 0.1 in X and Y each during the [4.5] observations.  We include a linear correction in our modeling (discussed in Section~\ref{sec:spot}) and find that slopes $-0.03$ in X and $-0.06$ in Y remove the Channel 1 phase effect. The Channel 2 phase effect is smaller and we fit it with slopes -0.05 in X and 0.01 in Y.  Figure~\ref{fig-spitzerkepler} shows the Spitzer photometry after pixel phase correction. The noise estimate from Poisson statistics is 0.45\% for [3.6] and 0.33\% for [4.5] for a single observation; we estimate that the noise in the binned data shown in Figure~\ref{fig-spitzerkepler} is 0.10\% at [3.6] and 0.08\% at [4.5].

\subsection{Ground-based Photometry}

\begin{figure}
\plotone{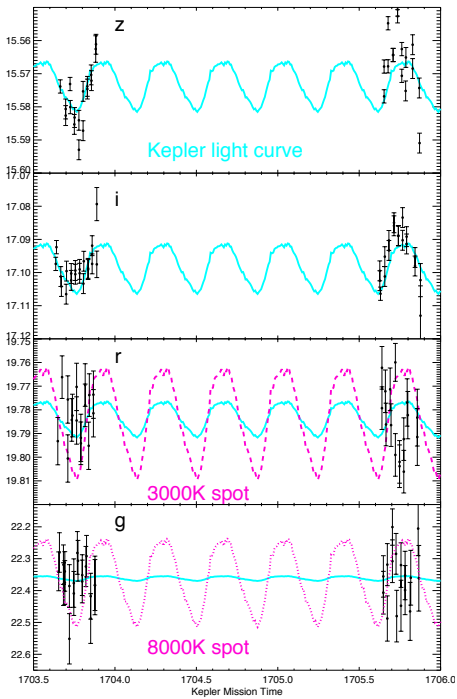}
\caption{Measured CCD photometry (in AB magnitudes) at $g$,$r$,$i$,$z$ compared to the average Kepler light curve extrapolated to the time of the Kitt Peak observations. Also shown is the predicted light curve at $r$ band if the Kepler light curve is due to a 3000K spot, and the predicted light curve at $g$ band if the Kepler light curve is due to an 8000K spot.
 \label{fig-kpno}}
\end{figure}

We observed W1906+40 using the Kitt Peak National Observatory (KPNO) 2.1-meter CCD direct imaging camera (Program 2013B-0340) on UT Dates 01 September 2013 and 03 September 2013, corresponding to Kepler mission dates 1703 and 1705; the other scheduled nights were too cloudy to observe.  We used the $g$ (KP1584), $r$ (KP1585), $i$ (KP1586), and $z$ (KP1587) filters with exposure times of 300, 300, 150 and 100 s respectively.  The W1906+40 aperture photometry was calibrated by comparison to background stars with Sloan Digital Sky Survey Data Release 9 measurements \citep{2012ApJS..203...21A}.  As an L1 dwarf, W1906+40 is much redder than the comparison stars, which has two consequences. First, there is differential extinction which we remove with a linear fit to airmass. Second, although our W1906+40 $g$ and $z$ magnitudes as calibrated by reference stars match the SDSS magnitudes, our $r$ and $i$ measurements are 0.4 and 0.2  magnitudes brighter. This is explained by the KPNO filters being wider in wavelength than the SDSS filters. The KPNO measurements are shown in Figure~\ref{fig-kpno} along with the predicted light curve based on extrapolating the Kepler average light curve.  The $i$ and $z$ observations are consistent with the expected Kepler light curve. The $g$ and $r$ curves are too noisy to detect the amplitude of the Kepler signal. 

W1906+40 was observed in the near-infrared (J band only) with the KPNO 2.1-meter on UT dates 24-28 October 2012 (Program 2012B-0233).  Because time-varying structures in the sky were evident, we chopped between two positions, taking five 60-second exposures at each position.  We used the alternate position to sky subtract and flat fielded with dome flats \citep{Joyce:1992fk}. We find, however, that the observed noise for W1906+40 and comparison stars is 2\% and we therefore did not detect the variations seen in the simultaneous Kepler data; the amplitude at J-band must be $<5\%$.  We make no further use of the J-band data.

\subsection{Ground-based Spectroscopy}

Optical (far-red) long-slit spectra were obtained with the MMT and Gemini-North telescopes to investigate whether chromospheric H$\alpha$ emission might track the Kepler optical variability. The MMT observations were on UT Date 26 and 27 August 2012 with the Red Channel spectrograph using grating 270, using 600 second exposures. Conditions were non-photometric. This time was allocated by NOAO as part of program 2012B-0233. The wavelength coverage was 6170 - 9810\AA~ with a resolution of $\sim12$\AA, but we make no use of the spectrum redder than 9250\AA~ due to strong telluric water absorption.  The Gemini-North observations (Gemini program GN-2012B-Q-105) were on UT Date  24 July 2012, 29 July 2012, and 04 August 2012 with the GMOS spectrograph \citep{Hook:2004lr} using grating R831 and 600 second exposures.  We thus have spectroscopy on Kepler mission dates 1299, 1304, 1314, 1332, and 1333.  
We processed the spectra with standard IRAF routines and measured the equivalent width using the task splot. T

\begin{figure*}
\plotone{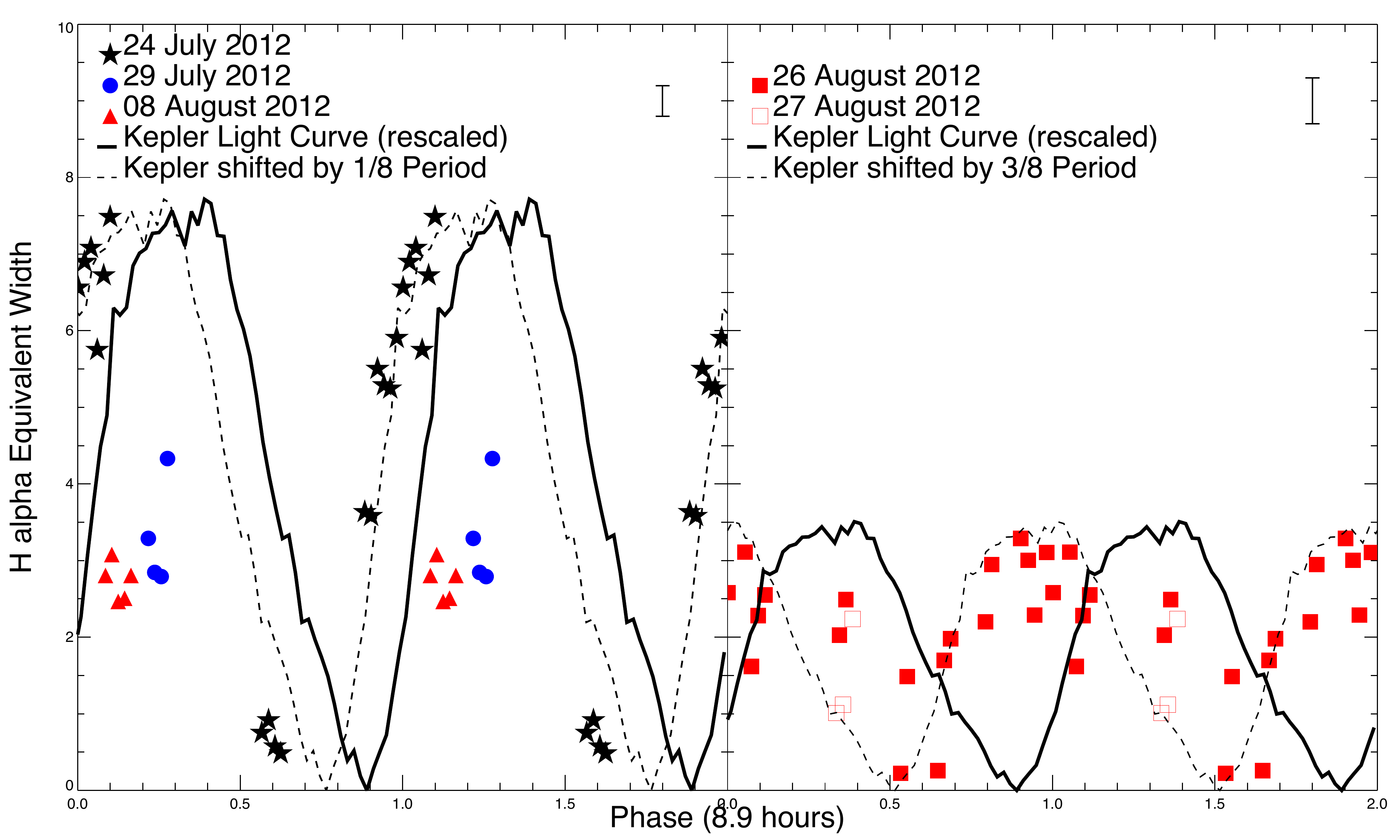}
\caption{Observed equivalent width of H$\alpha$ emission (in Angstroms) as a function of phase. The uncertainty for Gemini is $\pm 0.2$\AA~ and for MMT is $\pm 0.3$\AA; the size of these error bars is shown in the upper right of each panel. The systematically smaller MMT widths may be mainly due to the different instrumental resolutions (see text).  The Kepler light curve, arbitrarily scaled, is shown to guide the eye, along with versions shifted in phase. All data are plotted twice. The H$\alpha$ emission is not in sync with the broad-band optical variability.  
\label{fig-halpha}}
\end{figure*}

The phased H$\alpha$ data are shown in Figure~\ref{fig-halpha}.  The 24 July data begin with four observations taken near phase 0.6 with very weak H$\alpha$ (Equivalent width 0.5\AA)  in emission.  Our observations were then interrupted for a Target of Opportunity, but resumed with twelve observations covering phase 0.9 to 1.1, during which the H$\alpha$ emission increased from 3.5 to 8\AA.  The other Gemini nights had limited phase coverage. The 29 July 2012 had four observations at phase 0.2-0.3 before becoming dominated by the white light flares discussed in Paper 1.  The 04 August 2012 observations consist of four exposures near phase 0.1.  The MMT 26 August observations began at phase 0.33 and continued until phase 0.10.  The 27 August observations were cut short by clouds and cover only phases 0.3 to 0.4.  he random uncertainty in the H$\alpha$ equivalent widths is 0.2 \AA, but the MMT observations were taken at lower spectral resolution which results in a higher pseudo-continuum and systematically lower equivalent widths. It is apparent from Figure~\ref{fig-halpha} that the H$\alpha$ emission does not follow the Kepler light curve.  The 24 July 2012 data could be consistent with an active region rotating into view, but shifted by $\sim 1/8$ of a period from the Kepler phasing; however, observations on other nights are not consistent with this phasing either. The 26 August data follows neither the Kepler nor the 24 July light curves; we include a light curve shifted by 3/8 period to illustrate this.  

In Paper I, we described infrared spectroscopy of W1906+40 using the Keck II NIRSPEC near-infrared echelle spectrograph \citep{McLean:2000lr}. The 11 August 2011 observations reported there had a radial velocity of $-22.5 \pm 0.5$.  We report that a second observation, analyzed in the same way, on 10 Sep 2011 was $-24.3 \pm 1.3$, consistent with no significant variation.

\section{Spot Characteristics}\label{sec:spot}

The Spitzer [3.6] and KPNO $i$/$z$ observations are consistent with the phasing of the Kepler data.  We therefore begin our discussion by making the fundamental assumptions that we can treat W1906+40 as stable over the entire time period to compare observations made at different times, and that the optical and mid-infrared variability is due to the same source, which for convenience we will term the ``spot" regardless of the physical cause and whether it is actually many features.

The long-term stability of the spot suggests to us that a simple geometric model is appropriate. In principle, any observed light curve can be matched by a sufficient number of spots, so we could fit the light curve with brighter (warmer) or darker (cooler) spots.  We observe that the average Kepler light curve and the Spitzer [3.6] curve both appear asymmetric in the sense that the brightest portion (phases 0.2 to 0.4 in Figures~\ref{fig-keplerhist} and~\ref{fig-spitzerkepler}) is flatter than the darkest portion (phases 0.6-1.0).\footnote{In Paper I, we argued the Kepler light curve had no flat portion, but the new analysis, the second year of Kepler photometry, and the Spitzer photometry show that the brighter phases are flatter than the darker phases.} We begin by fitting dark spot models and later argue why bright spots are unlikely.  

\begin{figure}
\includegraphics[width=0.5\textwidth]{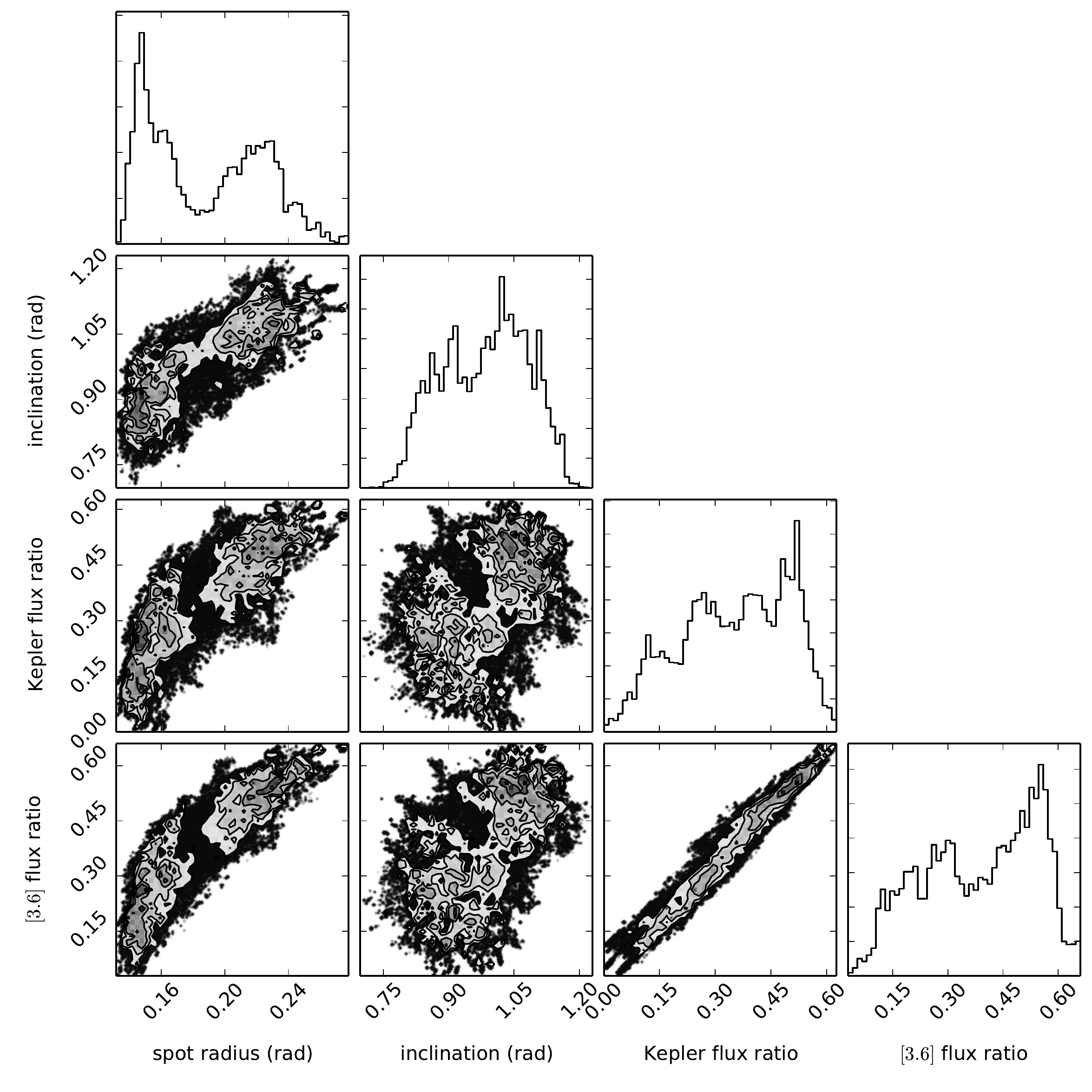}
\caption{Triangle plot of the projection of the multidimensional likelihood function
for the key single spot model parameters spot radius (radians), inclination (radians), Kepler flux ratio and Spitzer IRAC [3.6] flux ratio. These parameters are strongly correlated with each other. Other model parameters (Table~\ref{tab:parameters})  are not shown.
 \label{fig-triangle}}
\end{figure}

We therefore fit the Kepler and Spitzer datasets using a single spot that is dark in both the Kepler and [3.6] bands. As in Paper I, we model the light curve using the single circular spot equations of \citet{Dorren:1987fk} which we implemented in Python. We presented results for completely dark spots in Paper I. In this case, we use a spot with a single geometry (spot radius in radians, latitude, initial longitude, stellar inclination, period) but allow the unspotted photosphere brightness and spot-to-star flux ratio to vary for the three bands.  A list of all model parameters is given in Table~\ref{tab:parameters}.  
Here we define the spot-to-star ratio as the flux per unit area of the spot divided by the flux per unit area of the unspotted photosphere; this ratio is less than one for dark spots and greater than one for bright spots.  
We furthermore include X and Y linear pixel-phase corrections for the two IRAC bands. We adopt linear limb darkening parameters from \citet{2012A&A...546A..14C}.  For the photosphere, we use the 2300K model corresponding to an L1 dwarf: 0.84 for Kepler, 0.48 for [3.6], and 0.31 for [4.5].  For the spot, we adopt the 1900K limb darkening values, 0.84 for Kepler, 0.44 for [3.6] and 0.34 for [4.5]. We explore parameter space using an affine-invariant ensemble Markov chain Monte Carlo (MCMC) sampler \citep{Goodman:2010mc} as implemented in the Python code \texttt{emcee} \citep{2013PASP..125..306F} for maximum likelihood estimation. We use 200 walkers in the 11-dimensional parameter space to model the 23967 data points, and keep 1.41 million samples after discarding an initial burn-in sample.  The likelihood model is simply based on the $\chi^2$ statistic, i.e., $\ln p = -\frac{1}{2} \sum_n \left[ \frac{(f_n-m_n)^2}{s_n^2}\right]$ where for each of the $n$ data points, $f_n$ is the observed brightness, $s_n$ is the uncertainty, and $m_n$ is the model brightness given the trial parameters.  
The mean acceptance fraction was 0.16 and the mess auto-correlation length was 87. The period is $0.3701770 \pm 0.0000006$ days and the spot latitude is $1.21 \pm 0.06$ radians ($69.3 \pm 3.4$ degrees). The [4.5] star-to-spot flux ratio is $0.974 \pm 0.026$. The median spot size is 0.18 radians, but the spot size, other flux ratios and stellar inclination are highly correlated as we show in Figure~\ref{fig-triangle} with darker spots being smaller.  Median values and standard deviations for the fitted parameters are given in Table~\ref{tab:parameters}, but we caution that for the parameters shown in Figure~\ref{fig-triangle} the likelihood function is not a normal distribution. 

\begin{figure*}
\plotone{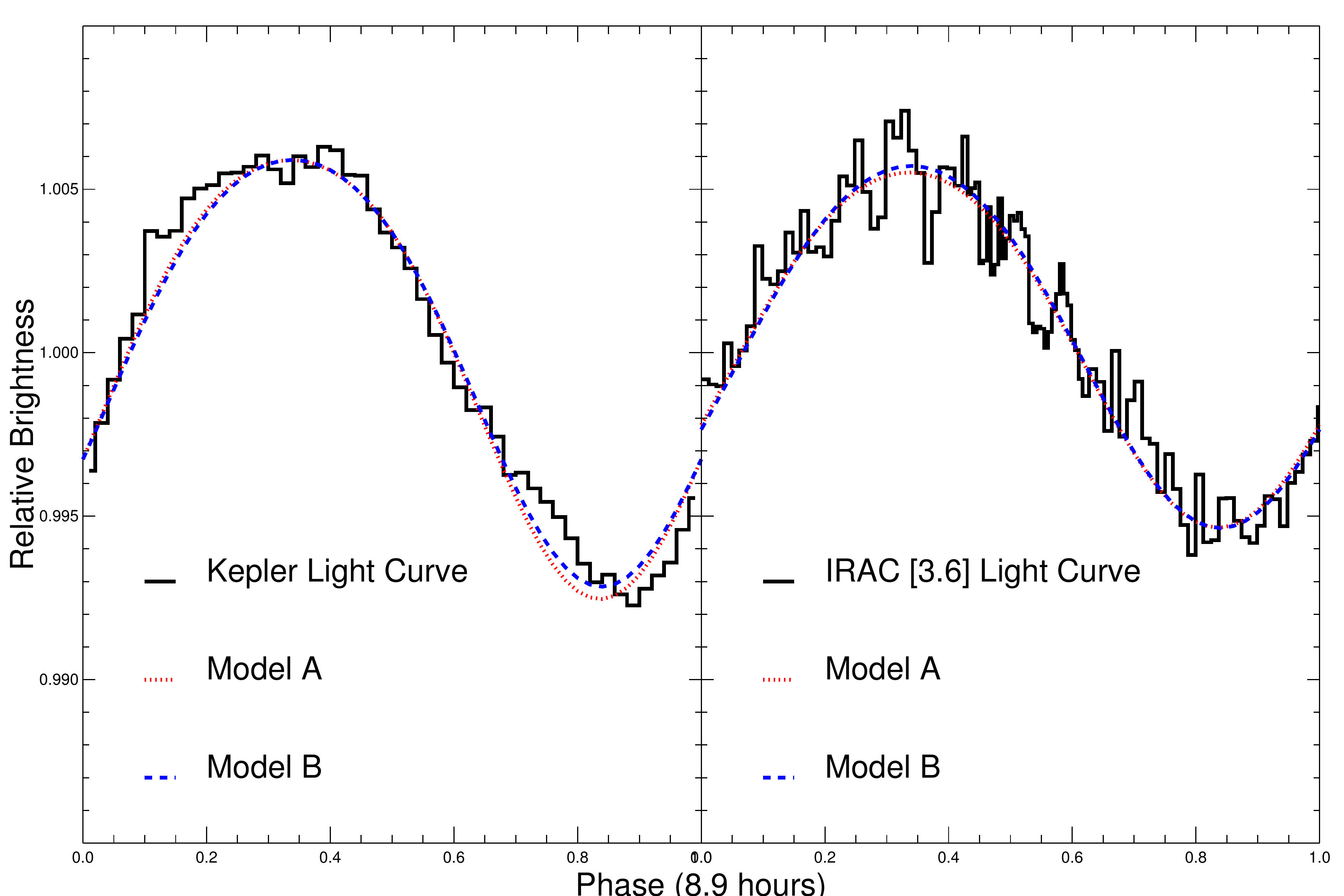}
\caption{Two sample single spot models shown with the Kepler and Spitzer [3.6] light curves.  Model A has spot radius 10.9$^{\circ}$, inclination 57.9$^{\circ}$, spot latitude 72.8$^{\circ}$, Kepler flux ratio 0.45, and IRAC [3.6] flux ratio 0.48. Model B has radius 9.2$^{\circ}$, inclination 56.7$^{\circ}$, spot latitude 74.5$^{\circ}$, Kepler flux ratio 0.20 and IRAC [3.6] flux ratio 0.20. We tested the goodness of fit with a k-sample Anderson-Darling test (included in the Python scipy package), and found that the null hypothesis that the models and averaged data are drawn from the same distribution cannot be rejected at even the 25\% significance level.  \label{fig-models}}
\end{figure*}

We show the light curve predicted by two of the models from the chain in Figure~\ref{fig-models}.  
Model A is typical of the solutions with a brighter spot: It has spot radius 0.190 radians, inclination 1.01 radians, spot latitude 1.27 radians, Kepler flux ratio 0.45, and IRAC [3.6] flux ratio 0.48. Model B is typical of the solutions with a darker spot; it has radius 0.16 radians, inclination 0.99, spot latitude 1.30 radians, Kepler flux ratio 0.20 and IRAC [3.6] flux ratio 0.20. In Paper I, we used the observed period, $v \sin i$, luminosity and effective temperature to show that high inclinations are preferred, with $\sin i > 0.59$.  Although we did not include this prior constraint in the MCMC simulations, all of the inclinations in Figure~\ref{fig-triangle} are acceptable, adding to our confidence in the single spot model.  The most important implication of the simulations is that a wide range of flux ratios are acceptable, from completely dark flux ratios of 0 to flux ratios $\sim 0.6$ in the Kepler and [3.6] bands. The Kepler-to-[3.6] flux ratios are similar ($0.91 \pm 0.12$). We caution, however, that the MCMC statistics should be viewed as illustrative rather than definitive; changes in the assumed limb darkening, for example, could affect any of the quoted parameters. These models also give us our best estimates of the relative peak-to-peak amplitude of the variability: 1.4\% for Kepler, 1.1\% at [3.6], and only 0.04\% at [4.5]. 

Although these circular dark spot models explain the main features of the light curve, it would be possible to fit the data with a spot (or set of spots) brighter than the photosphere in the Kepler and [3.6] filter. However, a number of possibilities can be ruled out from the multi-wavelength data.  
First, the variability cannot be due to a hot spot ($\teff \ge 3000$K) with continuum emission because of the $g$ and $r$ photometry. The unspotted photosphere has an L1 spectral type, confirmed by both the spectra and the SDSS/KPNO $griz$ photometry. The total number of additional counts from the spot through the full Kepler filter down to 430 nm can only be 1.4\% of the total. Putting these counts at the bluer parts of the Kepler filter, where the L dwarf photosphere is intrinsically faint, would require greater variability amplitudes ($\gg 1.4$\%) in the $g$ and $r$ filters. As an example, if these additional counts were distributed like an 8000K blackbody, the amplitude of variability at $g$ would be 0.3 magnitudes and at $r$ would be 0.08 magnitude. These values are ruled out by the data shown in Figure~\ref{fig-kpno}.  Even a 3000K blackbody would require an amplitude of 0.047 magnitudes at $r$ which is inconsistent with the optical data.  Second, a hot spot with emission lines is ruled out because the H$\alpha$ Balmer line is not strong enough, and no other red emission lines are observed.  Extremely strong blue emission lines are ruled out by the $g$ and $r$ photometry in the same way as a hot continuum. Furthermore, the H$\alpha$ variability (Figure~\ref{fig-halpha}) that is observed is not synched with the Kepler light curve. Third, as \citet{Heinze:2013uq} argued for their Spitzer observations of an L3 dwarf, the spot cannot emit like a blackbody or a normal M or L dwarf, whether warmer or cooler than the unspotted photosphere, in the mid-infrared. This is because such emission, like the unspotted photosphere, would have near-zero colors in [3.6]$-$[4.5] and therefore [4.5] would also have to be variable. A cold ($\teff < 1300$K) T dwarf-like spectrum is also ruled out. While a T dwarf-like spot would indeed be brighter at [4.5] than the Kepler or [3.6] filters, relative to the unspotted photosphere, the spot would still be very dark at [4.5] and therefore the light curve would end up being similar to [3.6].  

Despite the success of the single circular dark spot modeling, the small mismatches between the Kepler average light curve, the Spitzer [3.6] light curve, and the model light curves indicate a break-down in the assumption of a single, uniform, unchanging circular spot. The minimum in the average Kepler light curve around phase 0.9 is not present in the [3.6] curve and is not reproduced in our models. We note, however, that this feature is not present in the first year (Figure~\ref{fig-spitzerkepler}) Kepler light curve, only the second year, so one simple explanation is that the spot changed again between the end of the Kepler observations and the Spitzer.  We conclude that small changes are occurring in the geometry of the spot on the timescale of months to years, but most of the light curve can be explain with a large, mainly unchanging feature. To be sure, the circular spot is an approximation, and we cannot rule out the possibility of multiple circular spots, asymmetric bands or other geometric shapes.

\begin{deluxetable*}{lccl}
\tablewidth{0pc}
\tabletypesize{\footnotesize}
\tablenum{2}
\tablecaption{Parameters in Circular Spot Fitting}
\tablehead{
\colhead{Parameter} &  
\colhead{Median} &
\colhead{Standard Deviation} &
\colhead{Comment}
}
\startdata
Period (days) & 0.3701770 & 0.0000006 & \\
Latitude (degrees) & 69.3 & 3.4 & \\
Inclination (degrees) & 74.8 & 2.8 & See Figure~\ref{fig-triangle} for correlations\\
Spot Radius (degrees) & 10.5 & 2.1 &  See Figure~\ref{fig-triangle} for correlations\\
Spot-to-Star Ratio, Kepler & 0.37 & 0.14 & See Figure~\ref{fig-triangle} for correlations\\
Spot-to-Star Ratio, [3.6] & 0.41 & 0.15 & See Figure~\ref{fig-triangle} for correlations\\
Spot-to-Star Ratio, [4.5] & 0.974 & 0.026 & \\
X Linear Phase Correction, [3.6] & -0.03 & 0.002 &\\
Y Linear Phase Correction, [3.6] & -0.06 & 0.001 &\\
X Linear Phase Correction, [4.5] & -0.05 & 0.005 & \\
Y Linear Phase Correction, [4.5] & 0.01 & 0.005 & \\
Longitude (degrees) & 58.4 & 0.4 & At Mission Time 1180.0 days\\
Unspotted flux, Kepler & 1.0100 & 0.0007 & Median of all data set to one\\
Unspotted flux, [3.6] & 1.0104 & 0.0008 & Median of all data set to one \\
Unspotted flux, [4.5] & 1.0009 & 0.0004 & Median of all data set to one\\
Star Linear Limb Darkening, Kepler & 0.84 & \nodata & Not Fit, Adopted from Claret et al. (2012) \\
Spot Linear Limb Darkening, Kepler & 0.84 &\nodata & Not Fit, Adopted from Claret et al. (2012) \\
Star Linear Limb Darkening, [3.6] & 0.48 &\nodata & Not Fit, Adopted from Claret et al. (2012) \\ 
Spot Linear Limb Darkening, [3.6] & 0.44 &\nodata & Not Fit, Adopted from Claret et al. (2012)\\
Star Linear Limb Darkening, [4.5] & 0.31 &\nodata & Not Fit, Adopted from Claret et al. (2012) \\ 
Spot Linear Limb Darkening, [4.5] & 0.34 &\nodata & Not Fit, Adopted from Claret et al. (2012)\\
\enddata
\label{tab:parameters}
\end{deluxetable*}

\section{Discussion}

\begin{figure*}
\plotone{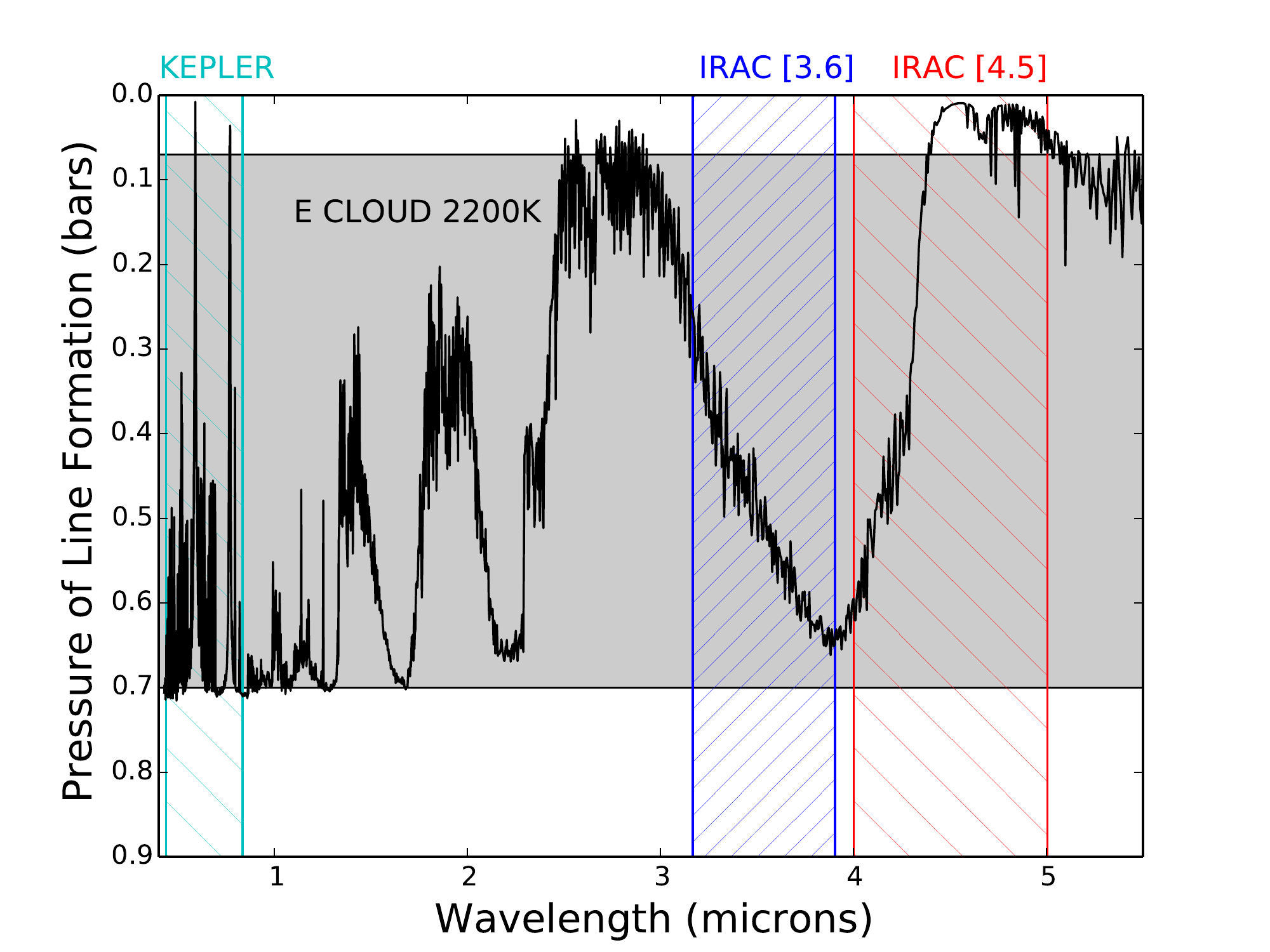}
\caption{Pressure of line formation ($\tau = 2/3$) in a 2200K ``E Cloud" model \citep{2001RvMP...73..719B,2006ApJ...640.1063B} compared to the positions of the Kepler and IRAC filters. The cloud layer is shown as shaded light grey. Due to the strong CO molecular opacity, IRAC [4.5] photometry probes low pressure (high altitude) while the Kepler and [3.6] photometry probe deeper layers.   
 \label{fig-pressure}}
\end{figure*}

Clouds provide a natural interpretation of our data, particularly the lack of variability at [4.5]. The pressure of line formation depth ($\tau = 2/3$) is shown in Figure~\ref{fig-pressure} for a 2200K, $\log g = 5.0$ ``E Cloud" model \citep{2001RvMP...73..719B,2006ApJ...640.1063B}.  The ``E cloud" provide the best description of L dwarfs in this model family, and describes a cloud deck which fully extends from 2300K (the base of the cloud, where calcium aluminates condense) up to the condensation
point of forsterite (1700-1900K), with exponential drop offs below and above these points. At [4.5], we only probe the outermost layers (lowest pressures and temperatures) of the atmosphere because of CO opacity, above the cloud tops (see Figure 9 of \citealt{2006ApJ...640.1063B}).  The [3.6] and Kepler bands, in contrast, match spectral windows where we probe deeper into the atmosphere, around the base of the cloud. Thus, as discussed by \citet{Gelino:2002uq}, a clearing in the cloud deck will cause brightening in the spectral windows; conversely, a thickening of the cloud deck will cause darkening. The dark spot implied by the single spot modeling would therefore correspond to a stable region of thicker cloud.  

Two solar system planetary analogs are of interest. For Jupiter, holes in the clouds produce bright spots in 5 \micron~images  \citep{1974ApJ...188L.111W}.  \citet{Gelino:2000lr} analyzed 5 \micron~data and showed that when analyzed as a point source, these complex, banded inhomogeneities produce a periodic rotational signal dominated by the Great Red Spot, which is {\it darker} at 5 \micron~ due to its thicker clouds.  Mark Marley (priv. comm, 2014) points to the example of Venus: In the near-infrared spectral windows such as 1.7 and 2.3 \micron~ where the thermal emission originates deep in the atmosphere, images of Venus's night side reveal cloud structures \citep{1984Natur.307..222A}; at other wavelengths, the night side of Venus appears uniform. Long-term monitoring shows that the clouds responsible for these structures, especially an equatorial band of variable thick clouds, change night-to-night, month-to-month, and year-to-year, with thicker-than-average clouds appearing darker and thinner-than-average clouds brighter \citep{2008P&SS...56.1435T}. Viewed as a point source, these varying clouds would produce photometric variability, but only at the wavelengths matching spectral windows.  Our observations are consistent with an analogous system; if we could spatially resolve W1906+40, the star would look uniform at [4.5] but cloud structures would be revealed in the optical and [3.6]. These clouds are non-uniform enough to produce the 1.4\% variability as the viewing angle changes due to the 8.9 hour rotation period.  Like the Great Red Spot but unlike Venus's clouds, W1906+40's cloud structure would be stable over several years.

For brown dwarfs, many groups (including \citealt{Rockenfeller:2006fk}, \citealt{Littlefair:2006fk}, \citealt{Radigan:2012vn}, \citealt{Heinze:2013uq}) have used one-dimensional model atmospheres to model patchy brown dwarfs, with the unspotted and spotted regions modeled with different effective temperature and cloud properties.  In the \citet{Ackerman:2001fj} approach, the $f_{sed}$ parameter controls the thickness of the clouds. \citet{Heinze:2013uq} found that an L3 dwarf had  [3.6] variability but none at [4.5]. They use the \citet{2008ApJ...689.1327S} family of models with this color constraint to find that either warmer, thinner or cooler, thicker clouds can produce this spectral signature. The same applies to the warmer W1906+40. Our preference for a darker spot because of the light curve suggests thicker clouds.  

Because the strongest evidence for clouds is the relative [3.6] and [4.5] photometry, we briefly discuss alternative models. One possibility is emission lines in the [3.6] band not present in the [4.5] band. An aurora model for variable [3.6] emission is discussed in detail by \citet{Heinze:2013uq}, who ultimately disfavor it for their L3 dwarf. This concept, discussed by \citet{Harding:2011fk} for optical emission in late-M dwarfs with periodic radio emission, incorporates a stable magnetic field with a beam of electrons and ions that excite the atmosphere, creating a bright spot. \citet{2000RvGeo..38..295B} and \citet{2014SSRv..tmp....7B} review aurorae emission mechanisms in the Solar System giant planets. The electrons ionize molecular hydrogen, leading to emission of photons from H$_3^+$ at  3-4\micron.  This would be responsible for the [3.6] emission; [4.5] would be constant due to the lack of molecular emission lines there. Visible emission at Kepler wavelengths is also expected. Through dissociative emission, the heating causes Balmer emission; other atomic and molecular emission \citep{1998JGR...10320113J} are also expected.  We disfavor this model for a number of reasons. First, a dark spot gives a better match to the light curve. Second, the H$\alpha$ data on different nights are not in phase with each other or with the Kepler light curve (Figure~\ref{fig-halpha}): The H$\alpha$ emission on 
26 August 2012 is bright when the Kepler light-curve is dim, yet on 24 July 2012 the H$\alpha$ is very weak when a Kepler bright spot would still be partially visible (phase 0.6) but then begins increasing (phase 0.9) before the bright spot rotates into view. Third, as discussed previously, continuum emission much bluer than the L dwarf photosphere would be inconsistent with the optical photometry. Finally,
we have no evidence of periodic radio emission (see Paper I), although the data do not include a full rotation cycle.  Another emission model would be a magnetically heated chromospheric active region. This is a likely explanation of the H$\alpha$ emission we do observe, and given the presence of strong white light flares, it is plausible that microflares or other processes heat regions of the chromosphere, but it cannot be responsible for the stable periodic Kepler light curve.  Finally, we note that  \citet{2014ApJ...785..158R} discuss thermal perturbations in brown dwarf atmospheres that might produce photometric variability on the timescale of hours or days as an alternative to changing cloud weather models. Along these lines, we might imagine a region of cooler gas at depths where $P \approx 0.5-0.7$ bar, producing the dark feature at Kepler and [3.6], without affecting the high altitude region measured by [4.5]. Unlike the thermal perturbations consider by \citet{2014ApJ...785..158R}, this cold gas would have to be stable for years, which would require an unknown mechanism to maintain.  

The most important qualitative difference from observations of cooler L, T, and Y dwarfs, especially the L/T transition, is the stability of the W1906+40 spot. There is strong evidence of night-to-night variations in many brown dwarfs suggesting rapid cloud evolution \citep{2013A&A...555L...5G}. In contrast, The W1906+40 light curve is stable over 843 days (over 2200 rotations); indeed, the standard deviation of the period in the MCMC simulations is just 50 milliseconds.  Evidently the W1906+40 spot is very long-lived compared to the ``weather" features in cooler L and T dwarfs.  Why would the clouds in W1906+40 be stable? \citet{2013ApJ...776...85S} and \citet{2014ApJ...788L...6Z} have presented exploratory models of the atmospheric circulation of brown dwarfs.  W1906+40's photosphere, however, is considerably hotter and is outside the range of these simulations. Future models may reveal whether  the W1906+40 spot could be an analog to the Great Red Spot.   As \citet{2013ApJ...779..101H} argue for the M9 dwarf TVLM 513-46546, whose optical variability is stable for 5 years, magnetic fields may play a role.  Perhaps a stable magnetic field suppresses convection to create a cool spot, which in turn develops thicker clouds. This mechanism would not function at later type L dwarfs, as shown by \cite{Gelino:2002uq}.  
Finally, we note that orbits have highly stable period. However, we have no evidence for a companion that could be affecting the W1906+40 photosphere.  

Our results demonstrate the importance of clouds even for the warmest L dwarfs. While ground-based studies of  $\sim 9$ hour period ultracool dwarfs will remain challenging, the Kepler K2 mission \citep{2014PASP..126..398H} is monitoring many more late-M and L dwarfs in campaigns of $\sim 75$ days. This will allow clouds and weather to be measured through the M/L transition and determine whether the stability of W1906+40's clouds is typical or unusual.  

\acknowledgments

We thank Adam Burrows and Mark Marley for discussion of their cloud models.  We are very grateful for the efforts of the Kepler, Spitzer, Kitt Peak National Observatory, Gemini and MMT teams who made these observations possible and responded so promptly and helpfully to questions.

This paper includes data collected by the Kepler mission. Funding for the Kepler mission is provided by the NASA Science Mission directorate. The material is based upon work supported by NASA under award No. NNX13AC18G.  This work is based in part on observations made with the Spitzer Space Telescope, which is operated by the Jet Propulsion Laboratory, California Institute of Technology under a contract with NASA. Support for this work was provided by NASA through an award issued by JPL/Caltech.
Some of the observations reported here were obtained at the MMT Observatory, a joint facility of the Smithsonian Institution and the University of Arizona. MMT telescope time was granted by NOAO, through the Telescope System Instrumentation Program (TSIP). TSIP is funded by NSF. This work is based in part on observations obtained at the Gemini Observatory, which is operated by the Association of Universities for Research in Astronomy (AURA) under a cooperative agreement with the NSF on behalf of the Gemini partnership: the National Science Foundation (United States), the Science and Technology Facilities Council (United Kingdom), the National Research Council (Canada), CONICYT (Chile), the Australian Research Council (Australia), CNPq (Brazil) and CONICET (Argentina). This research has made use of NASA's Astrophysics Data System, the VizieR catalogue access tool, CDS, Strasbourg, France, and the NASA/ IPAC Infrared Science Archive, which is operated by the Jet Propulsion Laboratory, California Institute of Technology, under contract with NASA. This work made use of PyKE \citep{2012ascl.soft08004S}, a software package for the reduction and analysis of Kepler data. This open source software project is developed and distributed by the NASA Kepler Guest Observer Office. This research has made use of IRAF, Astropy, a community-developed core Python package for Astronomy \citep{2013A&A...558A..33A}, and the triangle plotting code \citep{Foreman-Mackey:10598}. IRAF is distributed by the National Optical Astronomy Observatory, which is operated by the Association of Universities for Research in Astronomy (AURA) under cooperative agreement with the National Science Foundation.

\bibliographystyle{yahapj}
\bibliography{../astrobib}


\end{document}